\documentclass[12pt,draftcls,onecolumn,journal]{IEEEtran}

\usepackage{epsfig, epsf}
\usepackage{graphicx}
\usepackage{subfigure}
\usepackage[doublespacing]{setspace}
\usepackage[nospace,noadjust]{cite}
\usepackage{multirow}
\usepackage{array}
\usepackage{amsmath, amssymb}
\usepackage{amsthm}
\usepackage[ruled]{algorithm}
\usepackage{algpseudocode}
\usepackage{cite}
\usepackage{bm,upgreek}
\usepackage[T1]{fontenc}
\usepackage{etoolbox}

\newtheorem{remark}{Remark}

\begin{document}
\title{Optimal Fronthaul Compression for Synchronization in the Uplink of Cloud Radio Access Networks}
\author{Eunhye Heo, Osvaldo Simeone and Hyuncheol Park
\thanks{Eunhye Heo and Hyuncheol Park are with Department of Electrical Engineering, Korea
        Advanced Institute of Science and Technology (KAIST), Daejeon
        305-701, Korea (e-mail: hustery@kaist.ac.kr;
        parkhc@kaist.ac.kr).}
\thanks{Osvaldo Simeone is with Electrical and Computer Engineering Department, New Jersey Institute of Technology (NJIT), Newark, NJ, USA (e-mail: osvaldo.simeone@njit.edu).
The work of O. Simeone was partially supported by U.S. NSF under grant CCF-1525629.}}
\maketitle


\begin{abstract}
A key problem in the design of cloud radio access networks (CRANs)
is that of devising effective baseband compression strategies for
transmission on the fronthaul links connecting  a remote radio head (RRH)
to the managing central unit (CU). Most theoretical works on the subject implicitly
assume that the RRHs, and hence the CU, are able to perfectly recover time
synchronization from the baseband signals received in the uplink, and focus on the
compression of the data fields. This paper instead dose not assume a priori synchronization of RRHs and CU,
and considers the problem of fronthaul compression design at the RRHs with the aim
of enhancing the performance of time and phase synchronization at
the CU. The problem is tackled  by analyzing the impact of the
synchronization error on the performance of the link and by
adopting information and estimation-theoretic performance metrics
such as the rate-distortion function and the Cramer-Rao bound
(CRB). The proposed algorithm is based on  the Charnes-Cooper
transformation and on the Difference of Convex (DC) approach, and
is shown via numerical results to outperform conventional
solutions.
\end{abstract}

\begin{IEEEkeywords}
C-RAN, fronthaul compression, time and phase synchronization.
\end{IEEEkeywords}

\section{Introduction}
As mobile operators are faced with increasingly demanding
requirements in terms of data rates and operational costs,  the
novel architecture of cloud radio access networks (C-RANs) has
emerged as a promising solution \cite{CRAN1},\cite{CRAN2}. In a C-RAN, the
baseband processing of the base stations is migrated to a
central unit (CU) in the ``cloud", to which the base station, typically referred to an remote radio heads (RRHs), are connected
via fronthaul links, which in turn may be realized via fiber optics, microwave or mmwave technologies.
By simplifying the network edge and by centralizing baseband processing, the C-RAN architecture is expected to provide
significant benefits in energy efficiency, load balancing, and interference management capabilities (see review in \cite{CRAN2})

A key problem in C-RANs is that of devising
effective baseband compression methods in order to cope with the limitation
in the capacity of the fronthaul links. Most theoretical works on the subject
implicitly assume perfect time synchronization and channel state

information (CSI) at the RRHs and the CU (see, e.g.,
\cite{CRAN2}\cite{OS}). However, on the one hand, this
assumption violates the C-RAN paradigm that minimal baseband
processing should be carried out at the BSs, and, on the other
hand, the resulting design neglects the additional requirements
on fronthaul processing at the RRHs
that are imposed by synchronization and channel estimation. This
limitation is alleviated by \cite{PworkP}, which considers robust
compression in the presence of imperfect CSI, and by papers
\cite{Pwork4}\cite{PworkG}, which study the impact of fronthaul
compression on channel estimation. To the best of our knowledge, analysis that account for imperfect time synchronization are instead not available.

In this paper, we consider training-based synchronization for the
uplink of a C-RAN cellular system. Specifically, we investigate the problem of optimal fronthaul
compression of the training field with the aim of enhancing
the performance of time and phase synchronization at the CU.
To this end, the effect of the synchronization error on the signal to noise ratio (SNR) is analyzed by
adopting the Cramer-Rao bound (CRB) as the performance criterion of
interest and by accounting for compression via information theoretic tools. The resulting proposed algorithm is based on
the Charnes-Cooper transformation \cite{CCT} and the Difference of Convex (DC) approach \cite{DC}.  Numerical results show that
optimized fronthaul compression that targets enhanced synchronization performance outperforms conventional solution that do not account for the impact of synchronization errors.
The rest of the paper is organized as follows. Sec. II introduce system model of uplink C-RAN cellular system.
The analytic study of the performance and optimization are presented in Sec. III: the CRBs of time and phase offset estimation carried at CU is derived in Sec. III-A,
and the analysis of impact of the synchronization error on the effective SNR in Sec. III-B, and the optimization of fronthaul compression in Sec. III-C.
Finally, the performance is evaluated through simulations to present benefits of the proposed compression scheme in Sec. IV.

\section{System model}

In this paper, we consider training-based synchronization for the uplink of a
C-RAN cellular system. We specifically focus on the operation of a
single cell, as illustrated in Fig. 1, and assume that, as in
current cellular implementations, the MSs transmit over orthogonal
time/frequency resources, so that we can focus on a single active
MS in a given resource block. The MS transmits a frame consisting
of a training and a data field.  We further assume that the active
MS and the RRH have a single antenna. The RRH is connected to a CU
via a fronthaul link that can deliver $C$ bits per uplink sample
to the CU. It is also assumed that the RRH is synchronized at the
frame level so as to be able to distinguish between the training
and data fields that compose each transmitted frame.

\begin{figure} [!t]
\centering
\includegraphics[width=0.8\textwidth]{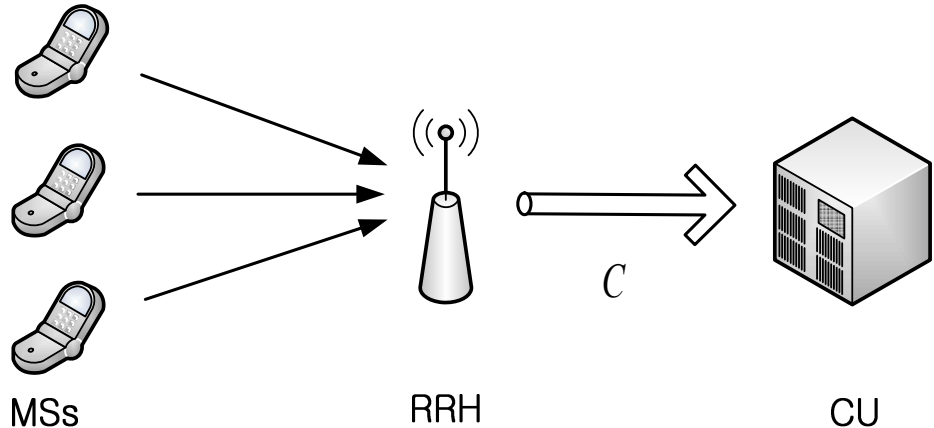}
\caption{Uplink communication between a number of MSs and an RRH.
The RRH is connected via a finite-capacity fronthaul link to a CU
that performs baseband processing, including synchronization. }
\label{system}
\end{figure}

\subsection{Training Phase}

Assuming a flat-fading channel, the signal received at the RRH during the training, or pilot, field,  is given as
\begin{align}
y_p(t) &= Ae^{j\theta}\sum _{ l=-L+1}^{N_p-1} x_p[l]g(t-lT-\tau) +
z_p(t), \ t\in [0,N_pT), \label{y_p}
\end{align}
where $A$ is a positive amplitude that accounts for the
attenuation due to fading; $\theta$ is the phase offset, which
models the effect of the channel and of the phase mismatch between
the oscillators at the MS and at the RRH; $\tau$ accounts for the
residual timing offset between MS and RRH; $T$ is the symbol
period; $x_p[l]$ is the $l$th pilot symbol transmitted by the MS;
$N_p$ is the number of pilot symbols; $g(t)$ is the pulse shape,
which includes the effect of the transmit and receive filter and
is assumed to be supported in the interval $[0,(L-1)T]$ for some
integer $L>1$; and $z_p(t)$ is the complex additive white Gaussian
noise with two-sided power spectral density $N_0$. We assume that
the RRH is able to estimate the channel amplitude $A$, for
instance,  by means of automatic gain control in the presence of
constant amplitude symbols. Instead, the time offset $\tau$ and
phase offset $\theta$ need to be estimated from the received
signal \eqref{y_p}.

The training sequence is generated randomly such that the symbols
$x_p[l]$  for $l=0,...,N_{p}-1$ are independent and distributed as
$\mathcal{CN}(0,E_{x_p})$. The training sequence is known to the
CU and the random generation is assumed here for the sake of
simplifying the analysis in the spirit of Shannon's random coding
(see, e.g., \cite{EL}). We further assume that the pilot symbols
are preceded by a cyclic prefix of duration equal to  $(L-1)T$.
This implies that $x_p[-l]= x_p[-l+N_p]$ for $1\leq l \leq L-1$.
Alternatively, as it will be discussed, the analysis below holds
as long as the number of training symbols $N_p$ is sufficiently
larger than  the support of the waveform $g(t)$ $L$.

In order to potentially enhance the performance of phase and time
synchronization, we allow the receiver to oversample the received
signal at the BS with a sampling period $T_s=T/F$, where $F$ is
the oversampling factor. We assume for simplicity of analysis that
a raised cosine pulse $g(t)$ with zero excess bandwidth (i.e., a
sinc function) is used, so that the two-sided bandwidth is
$B=1/T$. As a result, setting $F=1$, i.e., no oversampling, is an
acceptable choice that leads to no spectral aliasing. However, as it will be seen in Sec IV, the
selection $F>1$ may yield an improved performance. The resulting
discrete-time signal $y_p(mT+nT_s)$ can be expressed as the
interleaving of the $F$ polyphase sequences
$y^n_p[m]=y_p(mT+nT_s)$, with $n=0,1,...,F-1$, see, e.g.,
\cite{Shannon}. Each sequence $y^n_p[m]$ can be in turn written as
\begin{align}
y^n_p[m] &= A x_p[m]\circledast g_{\tau,\theta}^n[m] + z^n_p[m],\ \
m=0,...,N_p-1,  \label{sample_y}
\end{align}
where we have defined $z^n_p[m]\triangleq z_p(mT+nT_s)$,
$g_{\tau,\theta}^n[m]\triangleq e^{j\theta}g(mT+nT_s-\tau)$, and
$\circledast$ denotes the circular convolution. Assuming that the
noise $z_p(t)$ is white over the bandwidth $[-1/{2T_s},1/{2T_s}]$,
the discrete-time noise sequence $z^n_p[m]$ is an i.i.d. process
with zero mean and power $N_0/T_s$.

\begin{remark} Due to receive-side filtering, the noise is more properly
modelled as being bandlimited with the same bandwidth of the
signal. In this case, the discrete-time noise is actually
correlated across time for $F>1$. Here, following many
related references (see, e.g., \cite{whiteG2}\cite{whiteG3}), we
instead make the simplifying assumption that the noise is white.
This choice can be seen to lead to lower bounds on the actual
system performance. $\Box$
\end{remark}

\subsection{Data Phase}
The signal received during the data field of a frame can be written, in an analogous fashion as \eqref{y_p}, as
\begin{align}
y_d(t) &= Ae^{j\theta}\sum _{ l=-L+1}^{N_d-1} x_d[l]g(t-lT-\tau) +z_d(t), \ t\in [0,N_dT),
\end{align}
where $x_d[l]$ is the $l$th data symbol transmitted by the MS,
which is generated randomly in a constellation set $\Omega_x$ with
zero mean and power $E_{x_d}$, and $N_d$ is the number of data
symbols. The other parameters are defined as in \eqref{y_p}.
Moreover, as in \eqref{y_p}, we assume that the symbol indexed as
$l=-L+1,...,0$ amount to a cyclic prefix,  or that $N_p$ is
sufficiently larger than $L$. After sampling at baud rate for the
data field, the discrete-time signal is given as
\begin{align}
y_d[m] &= Ae^{j\theta}\sum _{ l=-L+1}^{N_d-1} x_d[l]g((m-l)T-\tau) +z_d[m], \ \ m=0,...,N_d-1,  \label{y_d}
\end{align}
where the discrete-time noise sequence $z_d[m]$ is an i.i.d.
process with zero mean and power $N_0/T$. Note that oversampling
could be adopted also for the data field by following the some
model used for the training filed, but we do not further pursue
this here in order to focus on training for synchronization.

\subsection{Fronthaul Compression}
Following the C-RAN principle, compression is performed at the RRH
in order to convey the baseband signal over the limited-capacity
fronthaul link to the CU. For the training field, we assume the
use of block quantizers that compress each $n$th polyphase
sequence $y^n[m]$, with $n=0,...,F-1$, separately for transmission
over the fronthaul link. Note that, while joint compression of
these sequences generally leads to an improved compression
efficiency, here we adopt separate compression both for its lower
computation complexity and for its analytical tractability. Using
the standard additive quantization noise model, the resulting
compressed signal for each $n$th polyphase sequence can be written
as
\begin{align}
\hat{y}^n_p[m] &= y^n_p[m]+q^n_p[m],  \ \ \ m=0,...,N_p-1, \label{y_com}
\end{align}
where $q^n_p[m]$ indicates the quantization noise. Noise
$q^n_p[m]$ is assumed, for simplicity of analysis, to be complex
Gaussian and generally correlated across the discrete-time index
$m$. From the covering lemma of rate-distortion theory \cite{EL},
vector quantization schemes can be designed such that the joint
(empirical) distribution of the input and output of the quantizer
satisfies \eqref{y_com}, as long as the rate is sufficiently
large (see, e.g., \cite{EL}). Furthermore, the relationship
\eqref{y_com} can be in practice approximated by a
high-dimensional dithered vector quantizers \cite{Zamir}.
The practical relevance of the additive-noise quantization model for system design
is further validated in Sec. IV by means of numerical results.

The covariance matrix $\textbf{K}_{\textbf{q}^n}$ of the vector
$\textbf{q}^n_p=[q^n_p[0],...,q^n_p[N_p-1]]$ is taken to be
circulant in order to facilitate its optimization in the frequency
domain as discussed in the next section. Due to the separate
quantization of the polyphase sequences, the quantization noise is
independent across the index $n$.

Taking the Discrete Fourier Transform (DFT) of \eqref{y_com} leads
to the frequency-domain signals
\begin{align}
 \hat{Y}^n_p[k] &= A X_p[k]G^n_{\tau,\theta}[k]+ Z^n_p[k]+ Q^n_p[k],  \label{F_y} \ \ k=0,...,N_p-1,
 \end{align}
where  $X_p[k]$, $G^n_{\tau,\theta}[k]$, $Z^n_p[k]$, and
$Q^n_p[k]$ are obtained by taking the DFT of the sequences
$\{x_p[m]\}_{m=0}^{N_p-1}$,
$\{g^n_{\tau,\theta}[m]\}_{m=0}^{N_p-1}$,
$\{z^n_p[m]\}_{m=0}^{N_p-1}$, and $\{q^n_p[m]\}_{m=0}^{N_p-1}$,
respectively. Due to the lack of spectral aliasing afforded by the
chosen waveform and sampling frequency, we can write
$G^n_{\tau,\theta}[k]=G^n[k]e^{-j(2\pi \frac{k}{N_pT_s}
\tau-\theta)}$.\footnote{The more general case with spectral
aliasing could be handled by using the analysis in \cite{Shannon}
and is left as an open problem.}

From the mentioned covering lemma \cite{EL} (see also
\cite{Zamir}), the fronthaul rate required to convey the compressed
signals
$\hat{\textbf{y}}_p=[\hat{\textbf{y}}_p^0,...,\hat{\textbf{y}}_p^{F-1}]$,
where
$\hat{\textbf{y}}_p^{n}=[\hat{y}^n_p[0],...,\hat{y}^n_p[N_p-1]]$,
from the RRH to the CU is given by the mutual information
$I(\textbf{y}_p;\hat{\textbf{y}}_p)$, with vector $\textbf{y}_p$
being similarly defined. However, the mutual information
$I(\textbf{y}_p;\hat{\textbf{y}}_p)$ depends on the joint
distribution of $\textbf{y}_p$ and $\hat{\textbf{y}}_p$ and hence
on the timing offset $\tau$ and phase offset $\theta$, which are
not known at the RRH. Therefore, the
necessary rate of a worst-case estimate is $R_p=\sup_{\tau,\theta}I(\textbf{y}_p;\hat{\textbf{y}}_p)$.
It can be easily calculated that the mutual information is given
by
\begin{align}
   I(\textbf{y}_p;\hat{\textbf{y}}_p) = \sum^{F-1}_{n=0}
  \log_2\frac{|\textbf{K}_{\textbf{y}^n_p}+\textbf{K}_{\textbf{q}^n_p}|}{|\textbf{K}_{\textbf{q}^n_p}|},
   \label{R_1}
\end{align}
where $\textbf{y}^n_p=[y^n_p[0],...,y^n_p[N_p-1]]$.  Since the
covariance matrix of the quantization noise
$\textbf{K}_{\textbf{q}^n_p}$ is assumed to be circulant, by
leveraging Szeg$\ddot{\textrm{o}}$ theorem \cite{szego}, we can
write \eqref{R_1} as
\begin{align}
I(\textbf{y}_p;\hat{\textbf{y}}_p) = \sum_{n=0}^{F-1}
  \sum_{k=0}^{N_p-1}
\log_2 \left( 1+ \frac{E_{x_p}A^2 |G^n[k]|^2+ N_0/T_s}{S_{Q^n_p}[k]}
\right), \label{R_3}
\end{align}
where $S_{Q^n_p}[k]$, for $k=0,...,N_p-1$, indicate the eigenvalues
of the matrix $\textbf{K}_{\textbf{q}^n_p}$. We will refer to
$S_{Q^n_p}[k]$ as the power spectral density (PSD) of the
quantization noise $q^n_p[m]$. We observe that \eqref{R_3} does not
depend on $\theta$ and $\tau$. Therefore, the required fronthaul
rate $R_p$ is given by the right-hand side of \eqref{R_3}. We will
therefore impose the fronthaul capacity constraint as
\begin{align}
I(\textbf{y}_p;\hat{\textbf{y}}_p)\leq N_pC, \label{I1}
\end{align}
where $I(\textbf{y}_p;\hat{\textbf{y}}_p)$ is given in \eqref{R_3}.

The compressed data signal during the data field, similar to
\eqref{y_com}, can be written as
\begin{align}
\hat{y}_d[m] &= y_d[m]+ q_d[m],\ \ m=0,...,N_d-1 .
\end{align}
where $q_d[m]$ indicates the quantization noise, which is assumed
to be white Gaussian random variable with zero mean and variance
$\sigma^2_{q_d}$. We observe that an optimized correlation for the
quantization noise on the data phase could also be designed,
similar to \cite{PworkP}, but we leave this aspect to future work
in order to concentrate on training for synchronization. Furthermore, following the discussion above, the
fronthaul rate required to convey the compressed data signal
$\hat{\textbf{y}}_d=[\hat{y}_d[0],...,\hat{y}_d[N_d-1]]$, from the
RRH to the CU is given by
$R_d=\sup_{\tau,\theta}I(\textbf{y}_d;\hat{\textbf{y}}_d)$, with
vector $\textbf{y}_d$ being similarly defined, with
\begin{subequations}
\begin{align}
   I(\textbf{y}_d;\hat{\textbf{y}}_d) &= \log_2\frac{|\textbf{K}_{\textbf{y}_d}+\textbf{K}_{\textbf{q}_d}|}{|\textbf{K}_{\textbf{q}_d}|},\\
   &= \sum_{i=0}^{N_d-1}\log_2 \left( 1+ \frac{E_{x_d}A^2 |G[i]|^2+ N_0}{\sigma^2_{q_d}} \right),\label{Rd_2}
\end{align}
\end{subequations}
where \eqref{Rd_2} follows from Szeg$\ddot{\textrm{o}}$ theorem as
in \eqref{R_3} and the fronthaul capacity constraint of the data
phase is given as
\begin{align}
I(\textbf{y}_d;\hat{\textbf{y}}_d)\leq N_dC. \label{I2}
\end{align}

\section{Analysis and Optimization}
In this section, we analyze the performance of the C-RAN system
introduced above by accounting for the impact of imperfect synchronization, with the aim of enabling the optimization of
fronthual quantization. We will first discuss the performance of
time and phase synchronization at the CU in Section III-A. Then,
we study the impact of synchronization errors on the SNR
in Section III-B. Finally, we investigate the optimization of
fronthual compression in Section III-C.

\subsection{CRBs for Time and Phase Offset Estimation}

The CU estimates the time and phase offsets based on the
compressed pilot signals $\hat{\textbf{y}}_p$, producing the
estimates $\hat{\tau}(\hat{\textbf{y}}_p,\textbf{x}_p)$ and
$\hat{\theta}(\hat{\textbf{y}}_p,\textbf{x}_p)$.  The mean squared
errors (MSEs) of these estimates can be bounded by the corresponding CRBs
i.e., by the inequalities
$E_{\hat{\textbf{{y}}}_p,\textbf{x}_p}[(\hat{\tau}(\hat{\textbf{y}}_p,\textbf{x}_p)
- \tau)^2]\geq \textrm{CRB}_ {\tau}$ and
$E_{\hat{\textbf{{y}}}_p,\textbf{x}_p}[(\hat{\theta}(\hat{\textbf{y}}_p,\textbf{x}_p)
- \theta)^2]\geq \textrm{CRB}_ {\theta}$. Note that the mentioned estimates
depend on  both the training sequence $\textbf{x}_p$ and the
compressed received signal $\hat{\textbf{y}}_p$,  and that the
squared error is averaged over the joint distribution of
$\textbf{x}_p$ and $\hat{\textbf{y}}_p$. To evaluate the CRBs, we
assume that the relationship \eqref{y_com}-\eqref{F_y} is
satisfied for the given vector quantizer. This is done for the
sake of tractability and is motivated by the covering lemma and by
the results in \cite{Zamir} as discussed in the previous section.
The CRBs are given, respectively, as
\begin{align}
\textrm{CRB}_{\tau}&= \left(\left(\frac{2\pi}{N_pT_s}\right)^2
\sum_{n=0}^{F-1}\sum_{k=0}^{N_p-1}\frac{E_{x_p} A^2k^2|G^n[k]|^2}{\frac{N_0}{T_s}+S_{Q^n_p}[k]}\right)^{-1},\label{CRB_tau}
\end{align}
and
\begin{align}
 \textrm{CRB}_{\theta}&=
\left(\sum_{n=0}^{F-1}\sum_{k=0}^{N_p-1}\frac{E_{x_p}|A|^2|G^n[k]|^2}{\frac{N_0}{T_s}+S_{Q^n_p}[k]}\right)^{-1}.
\label{CRB_theta}
\end{align}
Given \eqref{y_com}-\eqref{F_y}, the derivation of \eqref{CRB_tau}-\eqref{CRB_theta} follows from
standard arguments, see, e.g., \cite{MCRB}. Note that the bounds
\eqref{CRB_tau} and \eqref{CRB_theta} do not depend on the phase
$\theta$ and delay $\tau$.

\subsection{Impact of the Synchronization Error on the SNR}
Having estimated the time and phase offsets $\hat{\tau}$ and
$\hat{\theta}$, the CU compensates for these offsets the received
signal  \eqref{y_d}, obtaining the discrete-time signal
\begin{align}
y_d[m] &= Ae^{j\Delta \theta}\sum _{ l=-L+1}^{N_d-1} x_d[l]g((m-l)T+\Delta\tau) +z_d[m], \ \ m=0,...,N_d-1, \label{y_d}
\end{align}
where $\Delta\tau=\hat{\tau}(\hat{\textbf{y}},\textbf{x}) -\tau$
and $\Delta\theta=\hat{\theta}(\hat{\textbf{y}},\textbf{x})
-\theta$ are the synchronization errors for timing  and phase, respectively. We note that compensation of the time offset
requires interpolation, which is possible given the lack of
spectral aliasing. Moreover, under the mentioned assumption on the
zero excess bandwidth waveform $g(t)$, the statistics of the
(white Gaussian) noise terms are unchanged by interpolation.

To account for the impact of the synchronization errors $\Delta\tau$ and $\Delta\theta$, we follow the
approach in \cite{linear_approx}, whereby the sinc waveform $g(t)$
is approximated by retaining only two sidelobes on either side.
Under this approximation, we can express \eqref{y_d} as
\begin{align}
y_d[m] =  Ax_d[m]g(\Delta\tau)+z_s[m]+ z_{isi}[m] + z_d[m], \label{y_d1}
\end{align}
where the terms in \eqref{y_d1} are detailed below. First, the term
$z_{s}[m]=Ax_d[m]g(\Delta\tau)(e^{j\Delta \theta}-1)$ indicates
additional noise caused by the estimation error of phase offset
$\Delta \theta$. The term $z_{isi}[m]$ instead accounts for inter-symbol interference caused by the time synchronization error and is given as
\begin{align}
z_{isi}[m] = Ae^{j\Delta \theta}\sum_{l=m-3, l\neq m
}^{l=m+3}x_d[l]g((l-m)T+\Delta\tau). \label{y_isi}
\end{align}
In order to evaluate the power of the noise terms $z_{s}[m]$ and $z_{isi}[m]$,
we make the simplifying assumption that the estimation
errors $\Delta\tau$ and $\Delta\theta$ are uniform distributed on
$[-\frac{\Delta \tau_{\textrm{max}}}{2},\frac{\Delta
\tau_{\textrm{max}}}{2}]$ and on $[-\frac{\Delta
\theta_{\textrm{max}}}{2},\frac{\Delta
\theta_{\textrm{max}}}{2}]$, respectively. We observe that this
approximation is expected to be increasingly accurate in the
regime of small synchronization errors.  Moreover, we approximate
$\Delta\tau_{\textrm{max}}$ and $\Delta\theta_{\textrm{max}}$ by
means of the $\textrm{CRB}_ {\tau}$ \eqref{CRB_tau} and
$\textrm{CRB}_ {\theta}$ \eqref{CRB_theta}, respectively, by
imposing the equalities $E[\Delta\tau^2] = \textrm{CRB}_ {\tau}$
and $E[\Delta\theta^2] = \textrm{CRB}_ {\theta}$, which yields $\Delta\tau_{\textrm{max}}= \sqrt{12\textrm{CRB}_ {\tau}}$ and $\Delta\theta_{\textrm{max}}= \sqrt{12\textrm{CRB}_ {\theta}}$.
Finally, we adopt the piecewise linear approximation of the raised cosine
pulse $g(t)$ proposed in \cite{linear_approx}, whereby  pulse
$g(t)$ can be written as
\begin{subequations}
\begin{align}
g((l-m)T+\Delta\tau) \approx \ &a_l \times \frac{\Delta\tau}{T}, \label{g_ap1}\\
\textrm{where} \ \ &a_l=a_l^+  \ \  \textrm{if} \ \Delta\tau >0\\
\textrm{and} \ \  &a_l=a_l^-  \ \ \textrm{if} \ \Delta\tau <0,
\end{align}
\end{subequations} for $l\neq m$ and
\begin{align}
g(\Delta\tau)\approx  \left(
1-\eta\frac{|\Delta\tau|}{T}\right),\label{g_ap2}
\end{align}
where we have defined $\eta=\frac{2T}{\Delta
\tau_{\textrm{max}}}(1-g(\Delta \tau_{\textrm{max}}/2T))$ and the
values of $a_l^+$ and $a_l^-$ are listed in Table I, in which we have
$c_1= \frac{2T}{\Delta \tau_{\textrm{max}}}g(1-\frac{\Delta
\tau_{\textrm{max}}}{2T})$, $c_2=\frac{2T}{\Delta
\tau_{\textrm{max}}}|g(1+ \frac{\Delta
\tau_{\textrm{max}}}{2T})|$, $c_3=\frac{2T}{\Delta
\tau_{\textrm{max}}}|g(2-\frac{\Delta \tau_{\textrm{max}}}{2T})|$,
$c_4=\frac{2T}{\Delta \tau_{\textrm{max}}}g(2+\frac{\Delta
\tau_{\textrm{max}}}{2T})$, and $c_5=\frac{2T}{\Delta
\tau_{\textrm{max}}}g(3-\frac{\Delta \tau_{\textrm{max}}}{2T})$
\cite{linear_approx}.

\begin{table}[t]
\caption{Coefficients in the piecewise linear approximation of the raised cosine pulse}
\centering
\begin{tabular}{ | c || c | c | c | c | c | c |}
\hline $  l$ & $m-3$ & $m-2$ & $m-1$ & $m+1$ & $m+2$ & $m+3$ \\
\hline $a^+_l$ & $0$ & $c_4$ & $-c_2$ & $c_1$ & $-c_3$ & $c_5$ \\
\hline $a^-_l$ & $-c_5$ & $c_3$ & $-c_1$ & $c_2$ & $-c_4$ & $0$
\\ \hline
\end{tabular}
\end{table}
To evaluate the effect of the synchronization error on the performance, we now calculate an effective signal to noise ratio that accounts for the presence of the estimation
error for time and phase offsets. By using the approximations
discussed above, the following
approximations are derived in the Appendix. The power of the
desired signal $s_d[m]=Ax_d[m]g(\Delta\tau)$ in \eqref{y_d1} is
approximated as
\begin{align}
\textbf{E}_{\Delta\tau,x_d}[|s_{d}[m]|^2] &\approx A^2E_{x_d}
\left( 1-\frac{\eta}{2T}\sqrt{12 \textrm{CRB}_{\tau}}  \right),
\label{power_sd}
\end{align}
where $\textbf{E}_{a}[f(a)]$ denote the expectation of parameter
$a$ of function $f(a)$; the  power of $z_s[m]$ in \eqref{y_d1} is
similarly approximated as
\begin{align}
\textbf{E}_{\Delta\tau, \Delta\theta, x_d}[|z_s[m]|^2] &\approx
A^2E_{x_d} \textrm{CRB}_{\theta} \left( 1-\frac{\eta}{2T}\sqrt{12
\textrm{CRB}_{\tau}}  \right), \label{power_ys}
\end{align}
and the  power of $z_{isi}[m]$ in \eqref{y_isi} as
\begin{align}
\textbf{E}_{\Delta\tau,\mathbf{\bar{x}}_d}[|z_{isi}[m]|^2]
&\approx \frac{A^2E_{x_d}\bar{a}}{T^2}
\textrm{CRB}_{\tau},\label{P_isi}
\end{align}
where $\bar{a}=\Sigma^{l=m+3}_{l=m-3,l\neq m}|a_l|^2 $ and
$\mathbf{\bar{x}}_d=[x_d[m-3]\ x_d[m-2]\ x_d[m-1]\ x_d[m+1]\
x_d[m+2]\ x_d[m+3]]^T$.

Using \eqref{power_sd}, \eqref{power_ys}, and \eqref{P_isi}, we
obtain the approximate effective SNR expression
\begin{subequations}
\begin{align}
\textrm{SNR}_{\textrm{eff}} &\approx \frac{ A^2E_{x_d} \left( 1-\frac{\eta}{2T}\sqrt{12 \textrm{CRB}_{\tau}}  \right) }{ A^2E_{x_d} \textrm{CRB}_{\theta} \left( 1-\frac{\eta}{2T}\sqrt{12 \textrm{CRB}_{\tau}}  \right) + \frac{A^2E_{x_d}\bar{a}}{ T^2} \textrm{CRB}_{\tau} + \sigma^2_{z_d}+ \sigma^2_{q_d}}  \label{a_SINR}\\
 &\approx \frac{ A^2E_{x_d}}{ A^2E_{x_d} \textrm{CRB}_{\theta} + \frac{A^2E_{x_d}\bar{a}}{ T^2} \textrm{CRB}_{\tau} + \sigma^2_{z_d}+ \sigma^2_{q_d}}, \label{ap_SINR}
\end{align}
\end{subequations}
where, for analytical tractability, we made the further
approximation $1-\frac{\eta}{2T}\sqrt{12 \textrm{CRB}_{\tau}}
\approx 1$.  We observe that the expression \eqref{ap_SINR}
captures the effect of time and phase errors by means of additional noise
terms in the denominator of the effective SNR. We remark that the
approximations made in deriving \eqref{ap_SINR} will be validated
in the numerical results by evaluating the performance of proposed
optimization schemes for fronthaul compression that are based on \eqref{ap_SINR} and discussed next.

\subsection{Optimization of Fronthaul Compression}
In the proposed design, we wish to maximize the effective SNR \eqref{ap_SINR} under
the constraints \eqref{I1} and \eqref{I2} on the fronthaul capacity, over the statistics of
the quantization noises, namely over the PSDs ${S_{Q^n_p}[k]}$
corresponding to the quantization of the training field and over the
variance of the quantization noise $\sigma^2_{q_d}$ for the data
field.  Accordingly, we have following optimization problem:
\begin{subequations} \label{Max_2}
\begin{align}
\underset{\{S_{Q^n_p}[k]\},\sigma^2_{q_d}}{\textrm{maximize}} \ \ \ & \textrm{SNR}_{\textrm{eff}} \label{Max_ob}\\
\textrm{s.t.} \ \ \ & \sum_{n=0}^{F-1}\sum_{k=0}^{N_p-1} \log_2\left( 1+ \frac{E_{x_p}A^2 |G^n[k]|^2+\frac{N_0}{T_s}}{S_{Q^n_p}[k]}\right)\leq N_p C, \label{Max_con1} \\
& \sum_{i=0}^{(N-N_p)-1}\log_2 \left( 1+ \frac{E_{x_d}A^2 |G[i]|^2+ N_0}{\sigma^2_{q_d}} \right)\leq (N-N_p) C, \label{Max_con2} \\
&S_{Q^n}[k]\geq 0,\ \ \  n=0,...,F-1,\ k=0,...,N_p-1, \\
&\sigma^2_{q_d} \geq 0, N_p \geq 0,
\end{align}
\end{subequations}
where constraints \eqref{Max_con1} and \eqref{Max_con2} correspond to \eqref{I1} and \eqref{I2}, respectively.

Towards solving problem \eqref{Max_2}, we first observe that the variance  $\sigma^2_{q_d}$ can be obtained, without loss
of optimality,  by imposing the equality in constraint \eqref{Max_con2}. This is because $\textrm{SNR}_{eff}$ is monotonically decreasing with respect to $\sigma^2_{q_d}$ while the left-hand side of \eqref{Max_2} is
monotonically decreasing in $\sigma^2_{q_d}$.
We then have the following equivalent problem
\begin{subequations} \label{Min_3}
\begin{align}
\underset{S_{Q^n_p}[k]}{\textrm{minimize}} \ \ \ & A^2E_{x_d} \textrm{CRB}_{\theta} + \frac{A^2E_{x_d}\bar{a}}{ T^2} \textrm{CRB}_{\tau} \label{Min_ob}\\
\textrm{s.t.} \ \ \ & \sum_{n=0}^{F-1}\sum_{k=0}^{N_p-1} \log_2
\left( 1+ \frac{E_{x_p}A^2 |G^n[k]|^2+\frac{N_0}{T_s}}{S_{Q^n_p}[k]}
\right)\leq N_p C, \label{Min_con1} \\
&S_{Q^n}[k]\geq 0, \ \  n=0,...,F-1,\ k=0,...,N_p-1,
\end{align}
\end{subequations}
where the objective function \eqref{Min_ob} can be rewritten,
using \eqref{CRB_tau} and \eqref{CRB_theta}, as
\begin{align}
\frac{A^2E_{x_d} }{\sum_{n=0}^{F-1}\sum_{k=0}^{N_p-1}\frac{E_{x_p}A^2|G^n[k]|^2}{\frac{N_0}{T_s}+S_{Q^n_p}[k]}} + \frac{A^2E_{x_d}\bar{a}/T^2}{\left(\frac{2\pi}{N_pT_s}\right)^2
\sum_{n=0}^{F-1}\sum_{k=0}^{N_p-1}\frac{E_{x_p}A^2k^2|G^n[k]|^2}{\frac{N_0}{T_s}+S_{Q^n_p}[k]}}.
\end{align}

To tackle the optimization problem \eqref{Min_3}, we first define the auxiliary variables
 $u_{n,k}\triangleq(S_{Q^n}[k])^{-1}$,
$a_{n,k}\triangleq\left(2\pi/(N_pT_s)\right)^2k^2$
$E_{x_p}|A|^2|G^n[k]|^2$, and $b_{n,k}\triangleq
E_{x_p}|A|^2|G^n[k]|^2$, and then use the Charnes-Cooper
transformation \cite{CCT}, i.e., we set $v_{n,k}=(1+(N_0/T_s)u_{n,k})^{-1}$,
yielding the equivalent objective function
\begin{align}
\frac{A^2E_{x_d}}{\sum_{n=0}^{F-1}\sum_{k=0}^{N_p-1}
\frac{a_{n,k}}{N_0/T_s}(1-v_{n,k})} +
\frac{A^2E_{x_d}\bar{a}/T^2}{\sum_{n=0}^{F-1}\sum_{k=0}^{N_p-1}
\frac{ b_{n,k}}{N_0/T_s}(1-v_{n,k})}. \label{f_ob}
\end{align}

 \begin{algorithm}
 \caption{ DC algorithm for  problem \eqref{Min_3}}\label{MM1}
 \begin{algorithmic}[1]
 \State Initialization: $i=0$ and $v_{n,k}^{(0)}=1$ for $n=0,...,F-1$, $k=0,...,N_p-1$
 \State Obtain $\{v_{n,k}^{(i+1)}\}_{n,k}$ as a solution of the following convex problem:
\begin{align} \label{MM1_LP}
\underset{v_{n,k}}{\textrm{minimize}} \ \ \ & \eqref{f_ob} \nonumber\\
\textrm{s.t.} \ \ \ &\sum_{n=0}^{F-1}\sum_{k=0}^{N_p-1} (e_{n,k}^{(i)}v_{n,k}+f_{n,k}^{(i)} -\log_2((N_0/T_s)v_{n,k}) \leq N_pC,\nonumber\\
& 0 \leq v_{n,k} \leq 1,\ \ \forall \ n,k
\end{align}
\State Set $i=i+1$ \State If a convergence criterion is satisfied,
stop;  otherwise, go to step 2. Return the obtained solution
$v_{n,k}^{(i)}$  for $n=0,...,F-1$, $k=0,1,...,N_p-1$.
\end{algorithmic}
\end{algorithm}
The objective function \eqref{f_ob} is convex with respect to
the variables $v_{n,k}$ since denominator of each term is an affine
 function of $v_{n,k}$, and the function $1/g(x)$ is convex if $g(x)$ is concave and
 positive. However, the constraint \eqref{Min_con1} is still not convex in the
variables $v_{n,k}$ for $n=0,...,F-1$, $k=0,...,N_p-1$.
Nevertheless,  it can be expressed as the sum of a concave and of
a convex function, i.e.,
\begin{align}
\sum_{n=0}^{F-1}\sum_{k=0}^{N_p-1} \Big(\log_2
\big(-b_{n,k}v_{n,k}+b_{n,k}+N_0/T_s\big)-\log_2\big((N_0/T_s)v_{n,k}\big)\Big) \leq N_pC. \label{li_con}
\end{align}
Therefore, the Difference of Convex (DC)  approach \cite{DC} can be leveraged to obtain an iterative
optimization algorithm. This is done by linearizing the concave part of
\eqref{li_con} at the current iterate $v_{n,k}^{(i)}$, where $i$
is the index of the current iteration, obtaining the locally tight
convex upper bound
\begin{align}
\log_2 (-b_{n,k}v_{n,k}+b_{n,k}+N_0/T_s)\leq
e_{n,k}^{(i)}v_{n,k}+f_{n,k}^{(i)}, \label{ub}
\end{align}
where
$e_{n,k}^{(i)}=-b_{n,k}/(\ln(2)(N_0/T_s+b_{n,k}-b_{n,k}v_{n,k}^{(i)}))$
and
$f_{n,k}^{(i)}=\log_2(-b_{n,k}v_{n,k}^{(i)}+b_{n,k}+N_0/T_s)-e_{n,k}^{(i)}v_{n,k}^{(i)}$.

The DC algorithm performs  successive optimizations of the
convex problem obtained by substituting the right-hand side of
\eqref{ub} for the concave part in \eqref{li_con} until convergence. Given the known
properties of the DC algorithm \cite{DC}, the proposed approach,
 summarized in Algorithm 1, provides a feasible solution at every iteration and converges to a local minimum of problem \eqref{Min_3}.
Moreover, since it only requires the solution of convex problems,
the algorithm has a polynomial complexity per iteration.

\section{Numerical Results}

In this section, we present numerical results to give insight into
optimal fronthaul compression for synchronization and to validate
the analysis presented in the previous sections. Throughout, we
set $A=0.7$ and the SNR during training phase and SNR during data
phase are defined as $\textrm{SNR}_p = E_{x_p}/(N_0/T_s)$ and
$\textrm{SNR}_d = E_{x_d}/(N_0/T)$, respectively.
\begin{figure}
\centering
\includegraphics[width=0.8\textwidth]{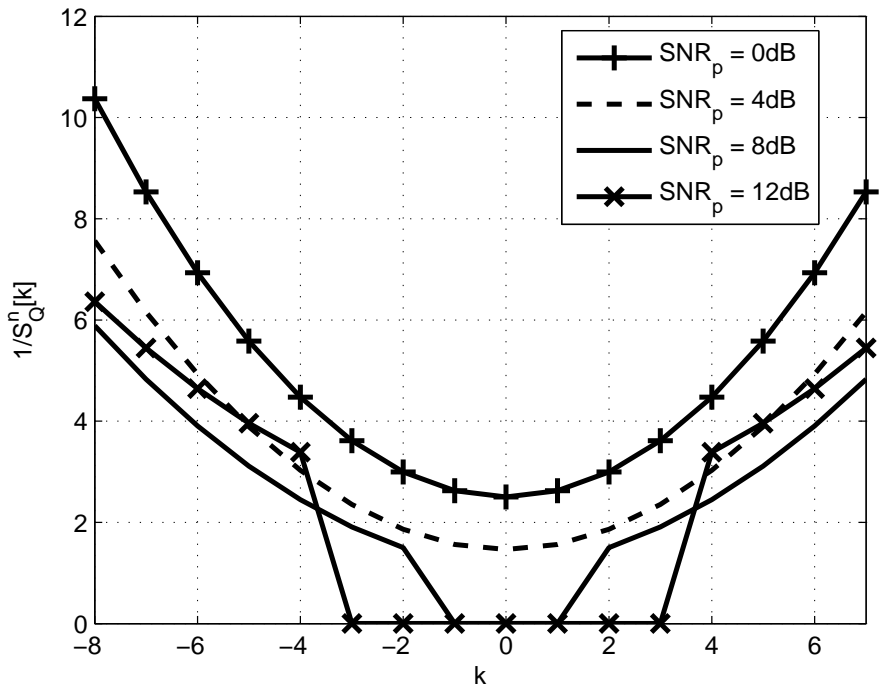}
\caption{Inverse of the PSD of the quantization noise obtained from Algorithm 1  versus the frequency index $k$ with $C = 3$ bits/sample,
$F=2$, $A=0.7$, $N=100$ and $N_p=16$.} \label{simul_optimalPSD}
\end{figure}

Fig. \ref{simul_optimalPSD} shows the inverse of the PSD of the
quantization noise $1/S_{Q^n_p}[k]$ obtained from Algorithm 1 for
various values of $\textrm{SNR}_p$ with $C=3$ bits/sample,
$N=100$, $N_p=16$, and $F=2$.  Note that the frequency axis ranges
from $-N_p/2$ to $N_p/2-1$ rather than in the interval $[0,N_p-1]$
for convenience of illustration. Moreover, we emphasize that
$1/S_{Q^n_p}[k]$is a measure of the accuracy of quantization at
frequency $k$ with $k = -N_p/2,..., N_p/2-1$,  so that a larger
$1/S_{Q^n_p}[k]$ implies a more refined quantization. We first observe
that the optimized solution prescribes a more accurate quantization at higher
frequencies, since these convey more information on the time delay,
as per the CRB \eqref{CRB_tau},  while all frequencies contribute in equal
manner to the estimate of the phase offset as per
\eqref{CRB_theta}. Moreover, as $\textrm{SNR}_p$ increases, it is
seen that lower frequencies tend to be neglected by the quantizer
in the sense that, for such frequencies, we have $1/S_{Q^n_p}[k]=0$, and
hence the signals on these frequencies are not compressed and not
transmitted to the CU.

In order to validate the advantage of the proposed design, we now
consider the synchronization performance under a conventional least-square joint phase and timing
estimator operating on the compressed signal $\hat{Y}^n[k],
n=0,...,F-1, k=0,...,N_p-1$. The estimator is given as
\begin{figure}
    \centering
    \subfigure[$F=1$, MSE of timing offset]
    {
        \includegraphics[width=0.5\columnwidth]{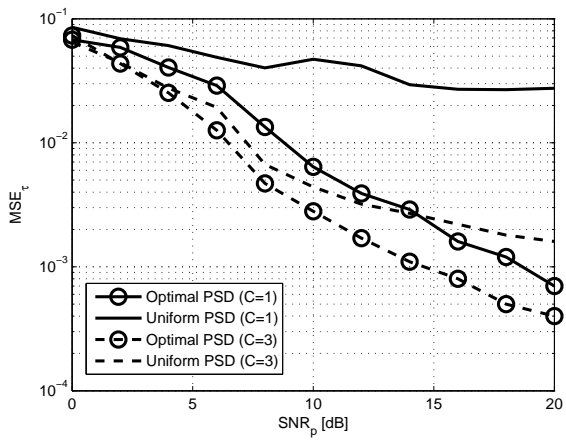}
        \label{fig:MSE_tau_F1}
     }\subfigure[$F=1$, MSE of phase offset]
    {
        \includegraphics[width=0.5\columnwidth]{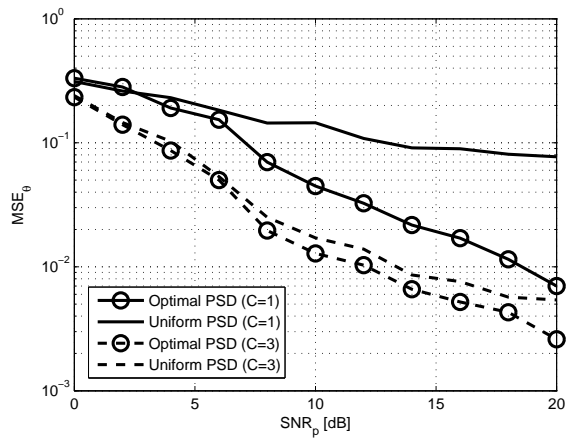}
        \label{fig:MSE_theta_F1}
    }

     \subfigure[$F=2$, MSE of timing offset]
    {
        \includegraphics[width=0.5\columnwidth]{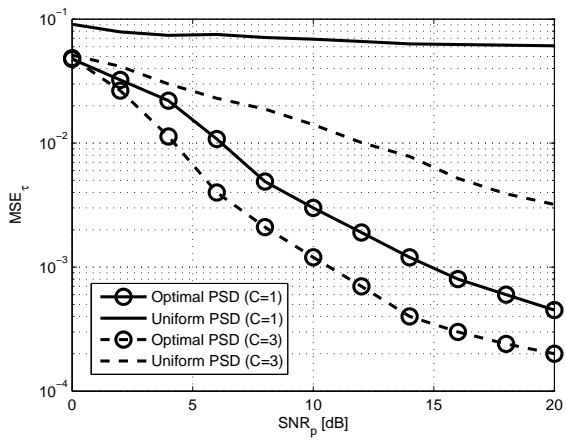}
        \label{fig:MSE_tau}
     }\subfigure[$F=2$, MSE of phase offset]
    {
        \includegraphics[width=0.5\columnwidth]{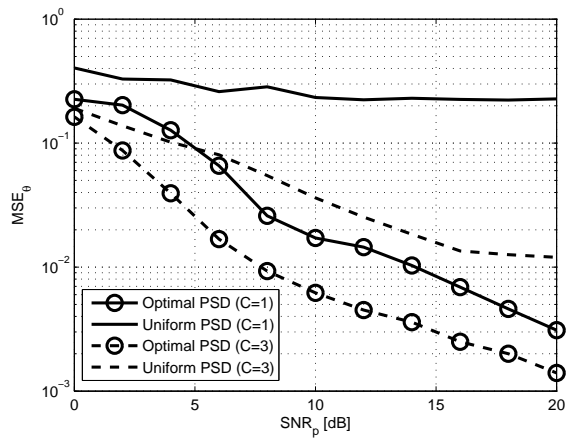}
        \label{fig:MSE_theta}
    }
    \caption{MSE for joint phase and timing estimation \eqref{joint_est} versus the $\textrm{SNR}_p$ with  $A=0.7$, $N=100$ and $N_p=16$.}
    \label{simul_MSE}
\end{figure}
\begin{align}
(\hat{\theta}, \hat{\tau}) = \arg \min_{\tilde{\theta},
\tilde{\tau}} \Phi(\tilde{\theta}, \tilde{\tau}),\label{joint_est}
\end{align}
with $\Phi(\tilde{\theta}, \tilde{\tau})=
\sum_{n,k}|r^n_k-r^n_k(\tilde{\theta}, \tilde{\tau})|^2$ where
$r^n_k=\arg(\hat{Y}^n[k]X^*[k])/{2\pi}$ and $r^n_k(\tilde{\theta},
\tilde{\tau})=\tilde{\theta}-k/N_p(n+\tilde{\tau})$. We evaluate the performance of the estimator \eqref{joint_est} in terms of MSEs of time and phase offsets by considering
the quantization noise with both the optimized PSD obtained from Algorithm 1 and a white PSD that is constant across all frequencies and is selected to satisfy the from the constraint \eqref{Max_con1}.
The white-PSD compression scheme is considered as reference as it does not attempt to optimize quantization with the aim of enhancing synchronization.

Fig. \ref{fig:MSE_tau_F1} and  Fig. \ref{fig:MSE_theta_F1} illustrate the
MSE of the timing and  phase offset estimates, respectively, as a
function of $\textrm{SNR}_p$ for $C = 1$ bits/sample and $C = 3$
bits/sample with $F=1$, $A=0.7$, $N=100$, and $N_p=16$. In addition,
we plot the MSE of the timing and phase offset estimates in case of $F=2$ in Fig. \ref{fig:MSE_tau} and  Fig. \ref{fig:MSE_theta}, respectively, under the same parameters.
We observe that the proposed scheme significantly outperforms the
conventional white-PSD strategy, and that the gain of the
proposed scheme is more pronounced for larger SNR values. This is
because as the SNR grows, the impact of the quantization noise
becomes more relevant compared to the channel noise. Furthermore, a larger oversampling factor $F$ is seem to yield an improved performance
only for the proposed optimization scheme and not with the conventional white-PSD scheme. This is because in the latter case,
the performance benefits of a larger number of observation are offset by the increased fronthaul overhead, which leads to a more pronounced quantization noise.
\begin{figure}
\centering
\includegraphics[width=0.8\columnwidth]{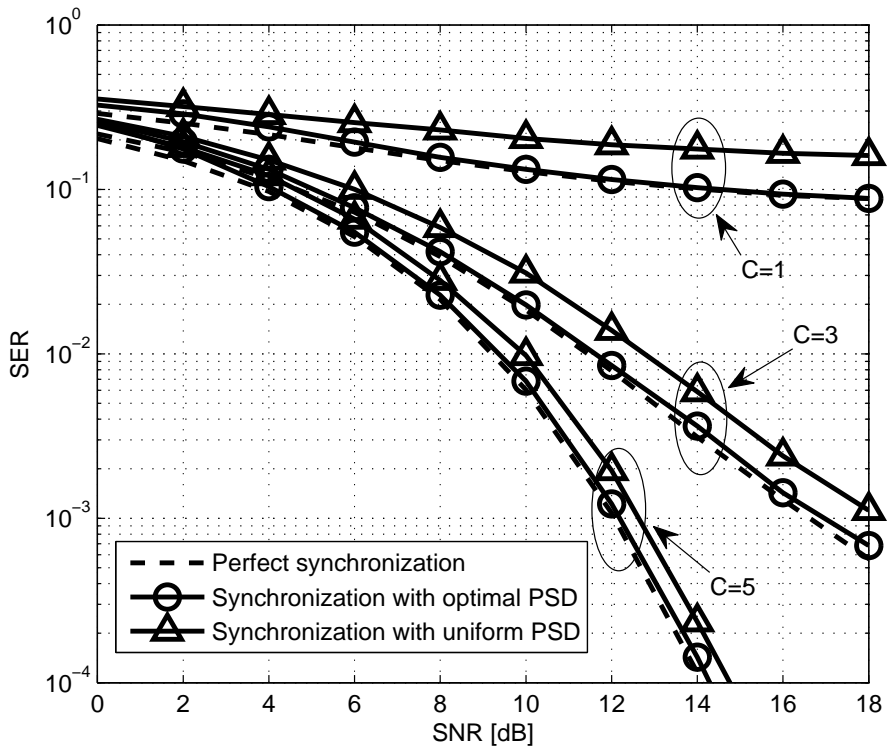}
\caption{SER with uncoded BPSK transmission versus SNR with joint phase and timing estimation \eqref{joint_est}, $F=2$, $A=0.7$, $N=100$ and $N_p=16$.} \label{simul_SER_BPSK}
\end{figure}
\begin{figure}
\centering
\includegraphics[width=0.8\columnwidth]{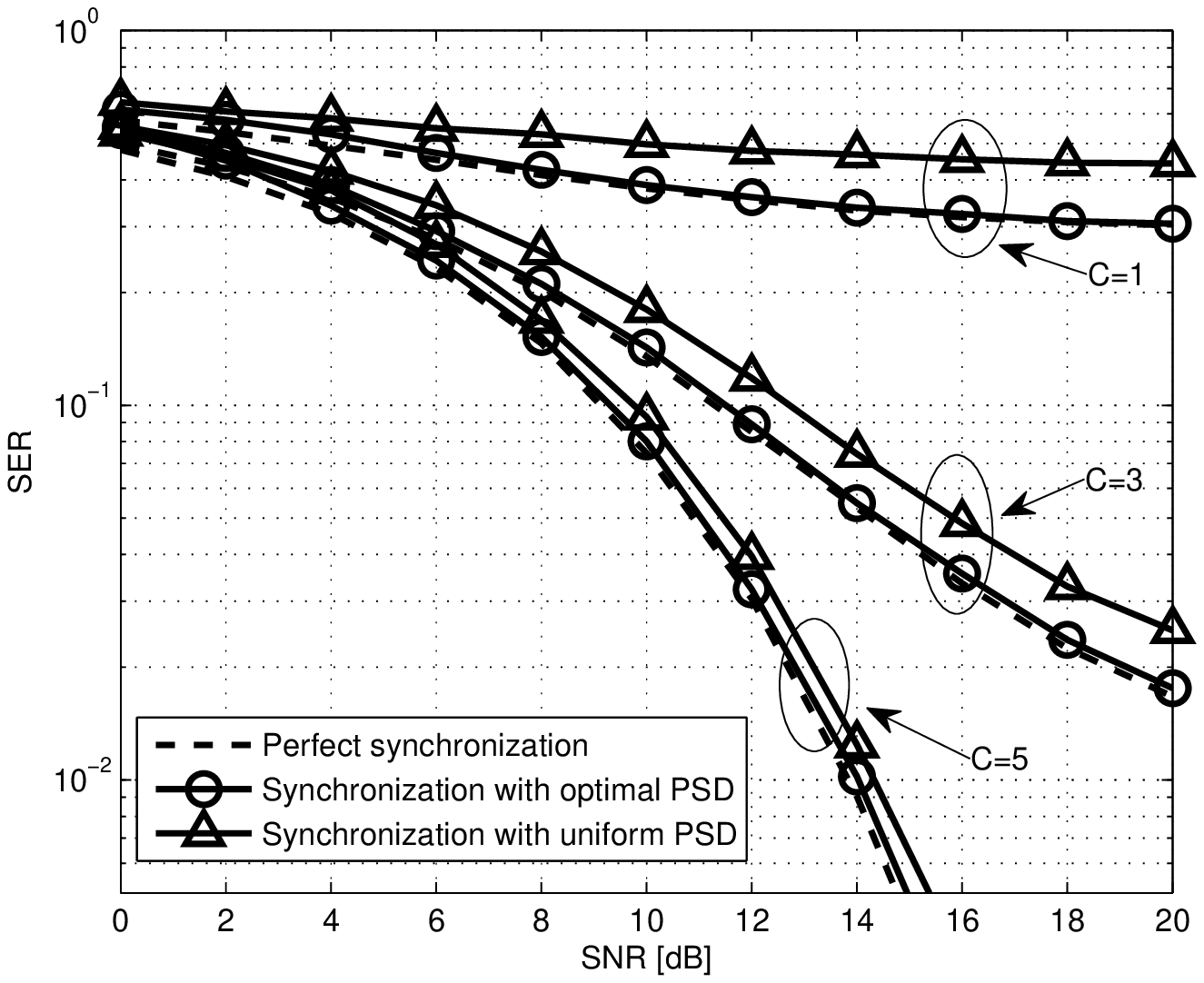}
\caption{SER with uncoded QPSK transmission versus SNR with  joint phase and timing estimation \eqref{joint_est},$F=2$, $A=0.7$, $N=100$ and $N_p=16$.} \label{simul_SER_QPSK}
\end{figure}
\begin{figure}
\centering
\includegraphics[width=0.8\columnwidth]{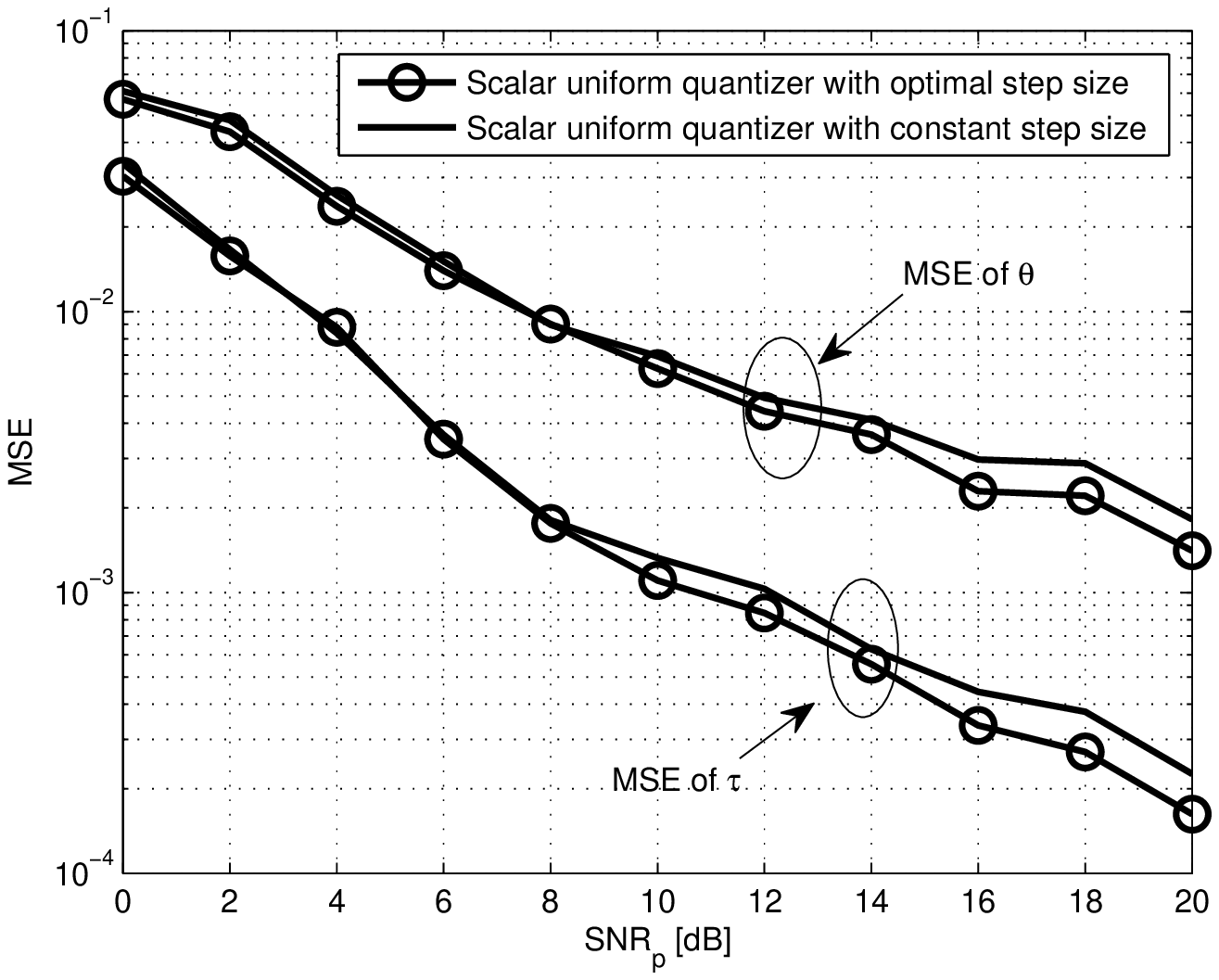}
\caption{MSE of joint phase and timing estimation versus $\textrm{SNR}_p$ in the presence of scalar fronthaul quantization and joint phase and timing estimation \eqref{joint_est} with $F=2$, $C=3$, $A=0.7$, $N=100$ and $N_p=16$.} \label{MSE_scalar_quan}
\end{figure}

Adopting the same estimator for time and phase offset, the system performance in terms of uncoded SER during the data
phase is shown in Fig. \ref{simul_SER_BPSK} and Fig.
\ref{simul_SER_QPSK} for BPSK and QPSK modulation, respectively,
versus the SNR for both training
and data fields, i.e., $\textrm{SNR} = \textrm{SNR}_p =
\textrm{SNR}_d$, with $F=2$, $A=0.7$, $N=100$ and $N_p=16$.
Simulation results with perfect synchronization are also presented for reference.
We note that, consistently with the
results in Fig. \ref{simul_SER_QPSK}, the proposed method is observed
to outperform  the conventional white-PSD scheme more
significantly as the SNR increases and as the fronthaul capacity
$C$ decreases. For instance, it is seen in
Fig. \ref{simul_SER_QPSK} that the proposed approach has a gain of
about 0.5 dB for $C=5$ bits/sample, of about 2 dB for $C=3$
bits/sample, and of about 10 dB for $C=1$ bits/sample at
sufficiently large SNR.

Finally, we elaborate on the performance of actual quantization by adopting a standard scalar uniform quantizer, instead of the additive quantization model considered so far.
In particular, we choose  the step size $\Delta[k]$ of the quantizer used for frequency $k$ based on the optimal PSD $S_q[k]$ obtained from Algorithm 1 by using the relationship $S_q[k]=\frac{|\Delta[k]|^2}{12}$.
This relationship is justified by fact that, at high resolution, the quantization noise is approximately uniformly distributed.
As reference, we also consider the performance of a uniform quantizer in which step size is same for all frequencies $k$, i.e., $\Delta[k]=\Delta$,
with the same dynamic range as for the optimized quantizer.
Fig. \ref{MSE_scalar_quan} presents the MSE of the timing and phase offset estimates versus $\textrm{SNR}_p$ with $F=2$, $C=3$, $A=0.7$, $N=100$ and $N_p=16$.
We observe that the proposed scheme outperforms the
conventional uniform quantizer, with a gain of about 2 dB in the high SNR regime.

\section{Conclusions}
This paper tackles the problem of optimal fronthaul compression
with the aim of enhancing the effective SNR in the presence of time and phase
synchronization errors at the CU. The proposed algorithm optimizes the
PSD of quantization noise at the RRHs by using the Charnes-Cooper transformation
and the DC approach, and is shown to outperform the conventional
solution that assumes an equal quantizer at all frequencies. Numerical results validate the analysis by evaluating the performance of the proposed design
under practical synchronization algorithms and with scalar quantization.
An interesting direction for future research is the consideration of
frequency-selective channels and of frequency synchronization.

\appendix
In this Appendix, we compute the powers of the desired signal
$s_{d}[m]$ and of the interference terms $z_s[m]$  and of
$z_{isi}[m]$ as defined in Section III-B.  The power of the desired
signal  is approximated, using \eqref{g_ap2}, as
\begin{subequations}
\begin{align}
\textbf{E}_{\Delta\tau,x_d}[|s_{d}[m]|^2] &\approx A^2 \textbf{E}_{\Delta\tau,x_d}\left[ |x_d[m]|^2\left( 1-\frac{\eta}{T}|\Delta\tau|\right)^2 \right] \\
   &= A^2E_{x_d} \left( 1-\frac{2\eta}{T}\textbf{E}[|\Delta\tau|] + \frac{\eta^2}{ T ^2}\textbf{E}[|\Delta\tau|^2]\right)\\
  &= A^2E_{x_d} \left( 1-\frac{\eta \Delta \tau_{\textrm{max}}}{2T} + \frac{\eta^2}{T ^2}\frac{\Delta \tau_{\textrm{max}}^2}{12}\right)\label{AE_tau1} \\
  &\approx A^2E_{x_d} \left( 1-\frac{\eta\Delta \tau_{\textrm{max}}}{2T}  \right) \label{Atight_LB}\\
  &\approx A^2E_{x_d} \left( 1-\frac{\eta}{2T}\sqrt{12 \textrm{CRB}_{\tau}}  \right), \label{Apower_sd}
\end{align}
\end{subequations}
where in \eqref{AE_tau1} we used the assumption
$\Delta\tau \sim \emph{U}[-\frac{\Delta
\tau_{\textrm{max}}}{2},\frac{\Delta \tau_{\textrm{max}}}{2}]$,
which  implies  $\textbf{E}[|\Delta\tau|]=\frac{\Delta
\tau_{\textrm{max}}}{4}$  and
$\textbf{E}[|\Delta\tau|^2]=\frac{\Delta
\tau_{\textrm{max}}^2}{12}$; \eqref{Atight_LB} follows by removing
higher-order terms in $\Delta \tau_{\textrm{max}}$ under the
assumption that $\Delta \tau_{\textrm{max}}$ is small enough; and
\eqref{Apower_sd} is a consequence of the approximation
$\textbf{E}[\Delta\tau^2] = \frac{\Delta\tau_{\textrm{max}}^2}{12}
\approx \textrm{CRB}_{\tau}$.

The  power of $z_s[m]$ is similarly approximated, using
\eqref{g_ap2}, as
\begin{subequations}
\begin{align}
\textbf{E}_{\Delta\tau, \Delta\theta, x_d}[|z_s[m]|^2] & \approx A^2 \textbf{E}_{\Delta\tau, \Delta\theta, x_d}\left[ |x_d[m]|^2|e^{-j\Delta\theta}-1|^2\left( 1-\frac{\eta}{T}|\Delta\tau|\right)^2 \right] \\
   &= A^2E_{x_d} \textbf{E}_{\Delta\tau,\Delta\theta }\left[|e^{-j\Delta\theta}-1|^2\left( 1-\frac{2\eta}{T}|\Delta\tau| + \frac{\eta^2}{T ^2}|\Delta\tau|^2\right)\right] \label{AE_the1}\\
  &\approx A^2E_{x_d}\textrm{CRB}_{\theta} \left( 1-\frac{2\eta}{T}\textbf{E}_{\Delta\tau}[|\Delta\tau|] + \frac{\eta^2}{T ^2}\textbf{E}_{\Delta\tau}
  [|\Delta\tau|^2]\right)\label{AE_the2} \\
  &\approx A^2E_{x_d} \textrm{CRB}_{\theta} \left( 1-\frac{\eta}{2T}\sqrt{12 \textrm{CRB}_{\tau}}  \right), \label{Apower_ys}
\end{align}
\end{subequations}
where the approximation in \eqref{AE_the1} follows as
\begin{subequations}
\begin{align}
\textbf{E}_{\Delta\theta}[|e^{-j\Delta\theta}-1|^2] & = 2-2\textbf{E}_{\Delta\theta}[\cos(\Delta\theta)]\\
&= 2-2\left(\frac{\sin(\Delta \theta_{\textrm{max}}/2)}{\Delta \theta_{\textrm{max}}/2}\right)\\
&\approx 2-2\left(1- \frac{(\Delta \theta_{\textrm{max}}/2)^2}{3!}\right) \label{Ataylor}\\
&=\frac{\Delta \phi^2}{12}\\
&\approx \textrm{CRB}_{\theta}, \label{Apower_dt}
\end{align}
\end{subequations}
where \eqref{Ataylor} follows from the Taylor series of the sinc
function up to the second order, and \eqref{Apower_dt} is a
consequence of the approximation $E[\Delta\theta^2] =
\frac{\Delta\theta_{\textrm{max}}^2}{12} \approx
\textrm{CRB}_{\theta}$.

Finally, using \eqref{g_ap1}, the power of $z_{isi}[m]$  is
approximated as
\begin{subequations}
\begin{align}
E_{\Delta\tau,\mathbf{\bar{x}}_d}[|z_{isi}[m]|^2] & \approx \frac{A^2}{ T^2}E_{\Delta\tau,\mathbf{\bar{x}}_d}\left[ |\textbf{a}^T\mathbf{\bar{x}}_d|^2\Delta\tau^2 \right]\\
   &= \frac{A^2E_{x_d}\bar{a}}{T^2}
   E_{\Delta\tau}[\Delta\tau^2]  \\
   &\approx \frac{A^2E_{x_d}\bar{a}}{T^2} \textrm{CRB}_{\tau}.\label{AP_isi}
\end{align}
\end{subequations}

\end{document}